\documentclass[aps,twocolumn,groupedaddress,amsmath,amssymb,superscriptaddress]{revtex4-1}
\usepackage{graphicx}  
\usepackage{dcolumn}   
\usepackage{bm}        
\usepackage{verbatim}   
\usepackage{braket}
\DeclareMathOperator{\Mod}{mod}
\usepackage{subfigure}
\usepackage{gensymb}
\usepackage{mathtools}

\usepackage[export]{adjustbox}

\usepackage{color}
\usepackage[percent]{overpic}

\usepackage{stackengine}

\begin{document}

\title{Pseudogap and Fermi surface in the presence of spin-vortex checkerboard for 1/8-doped lanthanum cuprates}
\author{Pavel E. Dolgirev}
\affiliation{
   Skolkovo Institute of Science and Technology,
   Skolkovo Innovation Center, 3 Nobel St., Moscow 143026, Russia
   }
   \affiliation{
   Moscow Institute of Physics and Technology,
   9 Institutskiy per., Dolgoprudny, Moscow Region, 141700, Russia
   }
\author{Boris V. Fine}
\affiliation{
   Skolkovo Institute of Science and Technology,
   Skolkovo Innovation Center, 3 Nobel St., Moscow 143026, Russia
   }
\affiliation{
   Institute for Theoretical Physics, University of Heidelberg,
   Philosophenweg 12, 69120 Heidelberg, Germany
   }

\date{\today}
\begin{abstract}
Lanthanum family of high-temperature cuprate superconductors is known to exhibit both spin and charge electronic modulations around doping level 1/8. We assume that these modulations have the character of two-dimensional spin-vortex checkerboard and investigate whether this assumption is consistent with the Fermi surface and the pseudogap measured by angle-resolved photo-emission spectroscopy. We also explore the possibility of observing quantum oscillations of transport coefficients in such a background. These investigations are based on a model of non-interacting spin-1/2 fermions hopping on a square lattice and coupled through spins to a magnetic field imitating spin-vortex checkerboard. The main results of this article include (i) calculation of Fermi surface containing Fermi arcs at the positions in the Brillouin zone largely consistent with experiments; (ii) identification of factors complicating the observations of quantum oscillations in the presence of spin modulations; and (iii) investigation of the symmetries of the resulting electronic energy bands, which, in particular, indicates that each band is double-degenerate and, in addition, has at least one conical point, where it touches another double-degenerate band. We discuss possible implications these cones may have for the transport properties and the pseudogap.
\end{abstract}

\maketitle

\section{Introduction}

Several families of high-temperature cuprate superconductors are known to exhibit spin and/or charge modulations~\cite{tranquada1995evidence,yamada1998doping,hoffman2002four,mcelroy2003relating,vershinin2004local,hanaguri2004checkerboard,abbamonte2005spatially,mcelroy2005coincidence,wise2008charge,da2014ubiquitous,comin2015symmetry}. Resolving the character of these modulations acquired new urgency in recent decade in the context of the efforts to reconcile the angle-resolved photo-emission spectroscopy (ARPES) experiments~\cite{valla2006ground,chang2008electronic, he2009energy, matt2015electron} with the measurements of quantum oscillations of various observables in response to magnetic field~\cite{doiron2007quantum, vignolle2008quantum,sebastian2009spin,sebastian2012quantum, barivsic2013universal, doiron2015evidence}.  ARPES experiments in underdoped (hole-doped) cuprates generically observe open-ended lines of the Fermi surface known as Fermi arcs and accompanied by angle-dependent pseudogap.  At the same time, the observations of quantum oscillations indicate the presence of small closed Fermi surfaces. This phenomenology hinted at the possibility that the Fermi arcs originate from closed Fermi surfaces in a smaller Brillouin zone (BZ) emerging as a result of some kind of periodically modulated background. Such interpretations based on one-dimensional stripe-like or two-dimensional checkerboard-like charge modulations have indeed been proposed~\cite{millis2007antiphase,chakravarty2008fermi,chen2008quantum,zabolotnyy2009evidence,yao2011fermi,harrison2011protected,allais2014connecting}. Spin modulations have mostly been omitted in these interpretations because of the absence of the experimental evidence of static spin response in YBa$_2$Cu$_3$O$_y$ (YBCO) and other cuprate families exhibiting quantum oscillations. 

The cuprate family that does exhibit both spin and charge modulations is lanthanum cuprates. A priori, one may expect that the presence of spin modulations does not change the situation qualitatively, and hence some sort of quantum oscillations would be present. Moreover, the experiments of Ref.~\cite{laliberte2011fermi} showed that one of the quantities exhibiting quantum oscillations in YBCO \cite{doiron2015evidence}, namely, Seebeck coefficient, exhibits the same overall trends in both La$_{1.8-x}$Eu$_{0.2}$Sr$_x$CuO$_4$ (Eu-LSCO) and YBCO as a function of both temperature and doping.  Yet, no experimental evidence of quantum oscillations has been reported so far for Eu-LSCO or other lanthanum cuprates. This may be due to the difficulty of producing sufficiently high-quality samples, but there might also be deeper reasons.

\begin{figure}
\centering
\includegraphics[scale=0.6]{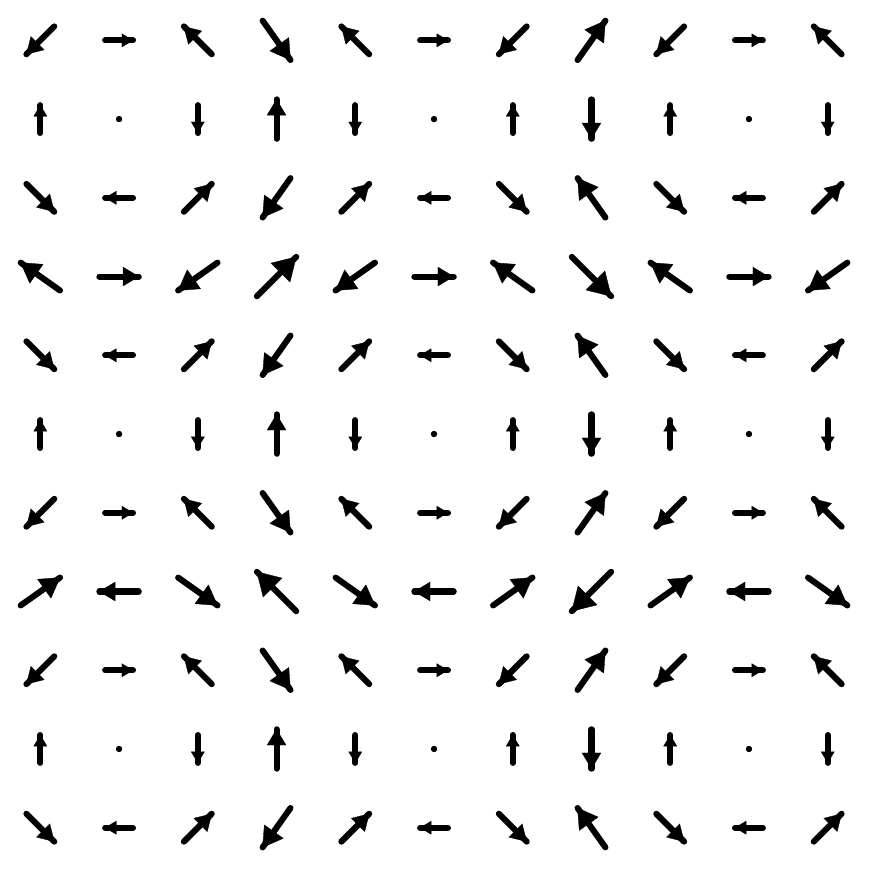}
\caption{Site-centered checkerboard corresponding to $\phi_1 = \phi_2 = 0$ in Eq.~(\ref{B})}
\label{fig: MLattice}
\end{figure}

The main focus of the present work is on 1/8-doped lanthanum cuprates, where elastic neutron scattering experiments~\cite{tranquada1995evidence} observe the four-fold splitting of the anti-ferromagnetic $(\pi,\pi)$ peak, and, at the same time, a later experiment~\cite{christensen2007nature} indicated that the modulation harmonics are linearly polarized in the direction transverse to the modulation wave vector. This leaves one with two possible interpretations, namely: (i) two domains of one-dimensional stripe-like modulations, or (ii) two-dimensional checkerboard of spin vortices shown in Fig.~\ref{fig: MLattice}. 
The above matter has been extensively discussed  on the basis of both theoretical arguments and experimental evidence~\cite{kivelson2003detect,fine2004hypothesis,Robertson2006,fine2007interpretation, fine2007magnetic, fine2011implications, brandenburg2013dimensionality}. On the theoretical side, the situation was, in particular, analysed on the basis of the Landau-type expansion in powers of the order parameter~\cite{Robertson2006}. This analysis indicated that the ground states of both stripe and checkerboard patterns are possible --- subject to material parameters, which are not known with the precision required to discriminate between the two possibilities. Microscopic models have also been investigated in this context --- see, e.g., Refs.~\cite{Zaanen1989,Seibold2011}, but here again, one can hardly rely on them, because they either neglect or very crudely approximate quantitatively important factors such as medium-range Coulomb interaction and/or electron-lattice coupling.  Various experiment-based arguments in favor of either stripes or checkerboards for lanthanum cuprates have been put forward in Refs.~\cite{kivelson2003detect,fine2004hypothesis,Robertson2006,SchriefferBook2007,fine2007interpretation, fine2007magnetic, fine2011implications, brandenburg2013dimensionality}, but the issue has not been settled either. This issue is elusive not only in lanthanum cuprates, but also for the yittrium-based and other cuprate families --- see, e.g., Refs.~\cite{comin2015broken,fine2016,Comin2016,Wang2015,Jang2016}. Recently, a somewhat similar situation emerged in the context of the ``spin-vortex crystal'' proposal for iron-based superconductors~\cite{Avci2014,Bohmer2015,Ohalloran2017,Meier2017}.

Fermi-surface reconstruction in the presence of stripe-like spin and charge modulations was described theoretically in Ref.~\cite{zabolotnyy2009evidence} in the context of ARPES experiments for 1/8-doped lanthanum cuprates.  Here our goal is to study the Fermi surface properties assuming the presence of the spin-vortex checkerboard shown in Fig.~\ref{fig: MLattice}. We develop a model of non-interacting fermions of spin-1/2 on a square lattice coupled through spins to local fields that mimic such a checkerboard.  These local fields originate from the exchange interaction, which, in turn, has its origin in the interplay of the Coulomb interaction and kinetic energy of electron. Therefore, in the leading order, the relativistic orbital effects of this local field can be neglected.

Our main results include: (i) the reproduction of Fermi arcs at the positions observed experimentally, (ii) the identification of factors complicating the observations of quantum oscillations in the presence of spin modulations, and (iii) the discovery that the model has a symmetry forcing each energy band to have at least one point, where it forms a conical connection to another band of the kind well-known from the physics of graphene~\cite{geim2007rise}. Such a property may drastically influence transport properties of the system. Moreover, this ubiquitous presence of conical points is a potential origin of the pseudogap. We also consider another scenario for the emergence of the pseudogap, which turned out to be more likely for the model parameters estimated to be relevant to lanthanum cuprates.

The article is organized as follows: In Section~\ref{sec: Model}, we formulate our model and discuss its relevance to 1/8-doped lanthanum cuprates. In Section~\ref{sec: Symmetry}, we investigate symmetries of the model and show that energy bands necessarily exhibit cones. In Section~\ref{sec: pseudogap}, we propose two scenarios for the emergence of the pseudogap, and, in Section~\ref{sec: Spin-vortex}, perform a calculation for a particular set of model parameters relevant to 1/8-doped lanthanum cuprates, thereby illustrating how our model can describe the pseudogap and the Fermi arcs. In Section~\ref{sec: Discussion}, we discuss various parameter regimes and possible generalizations of the model, and also place our results in the context of broader experimental knowledge about electronic transport in 1/8-doped lanthanum cuprates, addressing, in particular, the possibility to observe quantum oscillations. Finally, the main conclusions of the article are summarized in Section~\ref{sec: Conclusions}. 

\section{Model}
\label{sec: Model}

We consider a model of non-interacting spin-1/2 fermions on a square lattice in the background of $8\times 8$ periodic modulation of local fields as in Fig.~\ref{fig: MLattice}. The Hamiltonian is the following:
\begin{equation}
\mathcal{H} = \mathcal{H}_0 + \sum_{i,j;\alpha, \beta} a^{\dagger}_{ij,\alpha} \Big( B^x_{ij}S^x_{\alpha,\beta} + B^y_{ij}S^y_{\alpha,\beta}\Big) a_{ij,\beta}, 
\label{H}
\end{equation} 
where $i, j$ are the lattice indices; $\alpha, \beta$ are the indices of spin polarizations $\pm 1/2$ along the $z$-axis; $a_{ij,\alpha}$ are the fermionic annihilation operators; $S^x_{\alpha,\beta}$, $S^y_{\alpha,\beta}$ are the spin-1/2 operators;  $\mathcal{H}_0$ is the tight-binding Hamiltonian with hopping to the first, second and third nearest neighbors -- it has the following spectrum:
\begin{eqnarray}
E_0(\mathbf{k}) &=& 2  t  (\cos k_x  +\cos k_y) + 4  t' \cos k_x \cdot \cos k_y + \notag{}\\ 
&+& 2  t''  (\cos 2k_x + \cos 2k_y).
\label{E}
\end{eqnarray}
Magnetic field dependents on the lattice site position as follows:
\begin{eqnarray}
\mathbf{B}_{i,j} & = & (-1)^{i+j}\Big[ \begin{pmatrix}
B_0\\
0
\end{pmatrix} \cdot \cos\Big(\frac{\pi}{4}j + \phi_1\Big) +  \notag{} \\ 
& + &\begin{pmatrix}
0\\
B_0
\end{pmatrix} \cdot \cos\Big(\frac{\pi}{4}i + \phi_2 \Big)\Big], 
\label{B}
\end{eqnarray}
where $\phi_1$ and $\phi_2$ are two fixed phases of the two orthogonal harmonics. For Fig.~\ref{fig: MLattice}, $\phi_1 = \phi_2 = 0$, but we would like to consider the general case.

As the fermions fill one-particle states of the Hamiltonian $\mathcal{H}$, the system exhibits the $8 \times 8$  modulation of spin polarizations that follow the local magnetic field. It is accompanied by $4 \times 4$ checkerboard modulation of particle density of form $n_{i,j} = n_0 + \delta n_{i,j}$ with $\delta n_{i,j}  \propto |\mathbf{B}_{i,j}|^2$. These spin and charge densities modulations are consistent with the experiments~\cite{tranquada1995evidence}. This is, therefore, the minimal model describing the low-energy spin checkerboard response possibly emerging as a result of the delicate balance between large contributions from kinetic energy, Coulomb energy (including spin exchange) and electron-lattice interaction.

\subsection{Details of numerical solution}

We obtain density of states (DOS) $\nu(E)$ and other quantities of interest by directly diagonalizing Hamiltonian $\mathcal{H}$ in Eq.~(\ref{H}). We do it in the basis of the Bloch eigenstates of the Hamiltonian $\mathcal{H}_0$:  $\left\{ \Ket{\mathbf{k},\pm} \right\} $, where $+$ or $-$ represents the projection of particle's spin on the $z$-axis, and $\mathbf{k} \equiv (k_x, k_y)$ is a wave vector belonging to the the first Brillouin zone (BZ) of the square lattice: $-\pi \leq k_x, k_y < \pi$.  We refer to it as ``large BZ".

The $8 \times 8$ modulation of the local field reduces the BZ to $-\pi/8 \leq k_x, k_y < \pi/8$ (``small BZ''). The modulated terms in the Hamiltonian couple only those basis states in the large BZ which, after backfolding to the small BZ, have the same $\tilde{\mathbf{k}}$. These wave vectors are $\mathbf{k}_{l,m} = \tilde{\mathbf{k}} + \frac{\pi}{4}  l  \  \mathbf{e}_x + \frac{\pi}{4}  m \  \mathbf{e}_y$, where $\mathbf{e}_x \equiv (1,0)$, $\mathbf{e}_y \equiv (0,1)$, and  $l,m = 0,...7$.  Taking into account the fact that, for each of the 64 thus-defined wave vectors, there are also two spin states coupled by the local field terms, we obtain the energy spectrum for each $\tilde{\mathbf{k}}$ by diagonalizing the $128 \times 128$ Hamiltonian matrix, which has the following structure:
\begin{enumerate} 
\item $\Bra{\mathbf{k}_{l,m},\pm}\mathcal{H}\Ket{\mathbf{k}_{l,m},\pm} = E_0(\mathbf{k}_{l,m})$;
\item $\Bra{\mathbf{k}_{l_1,m_1},+}\mathcal{H}\Ket{\mathbf{k}_{l_2,m_2},-} =  \frac{i}{4}B_0 e^{\pm i \phi_2}$\\
for $(l_2 - l_1 \pm 1) [\Mod 8] = 4 $\\ and $(m_2 - m_1) [\Mod 8] = 4 $;
\item $\Bra{\mathbf{k}_{l_1,m_1},+}\mathcal{H}\Ket{\mathbf{k}_{l_2,m_2},-} = \frac{1}{4}B_0 e^{\pm i\phi_1}$\\
for $(m_2 - m_1 \pm 1) [\Mod 8] = 4$\\ and $(l_2 - l_1) [\Mod 8] = 4.$
\end{enumerate}
Elsewhere matrix elements are equal to zero.

\section{Symmetries and degeneracies of energy-bands}
\label{sec: Symmetry}

\subsection{Double degeneracy of energy bands}
We would like to show that each energy band in the spin-vortex
checkerboard model is twice degenerate.

Let $\hat{T}_x$ be an operator representing translation by 4 lattice
periods along the $x$-direction and subsequent rotation of spins through
$180^{\circ}$ about the $x$-axis. Analogously, let $\hat{T}_y$ be an
operator representing translation by 4 lattice periods along the
$y$-direction and subsequent rotation of spins through $180^{\circ}$
about the $y$-axis.  These operators have the following representation:
\begin{eqnarray}
&&\hat{T}_x \equiv \hat{\tau}_{(4,0)} \otimes (i\sigma_x)\\
&&\hat{T}_y \equiv \hat{\tau}_{(0,4)} \otimes (i\sigma_y)
\end{eqnarray}
where $\sigma_{\alpha}$  are the Pauli matrices, and
$\hat{\tau}_{\mathbf{a}}$ is translation by vector $\mathbf{a}$.

We now observe that operators $\hat{T}_x$ and $\hat{T}_y$ commute with
the Hamiltonian but do not commute with each other, and, in addition,
they do not change wave vector $\mathbf{k}$ of a fermionic state.
Therefore, each energy level for any given wave vector $\mathbf{k}$ is
at least twice degenerate, which means that each energy band is at least
double-degenerate.

\subsection{Conical touch points}

We now would like to show that at ${\bf k}_0 = (\frac{\pi}{8},
\frac{\pi}{8})$ each energy level is 4-times degenerate. This
wave vector  is a special high-symmetry point, because it is at the
corner of the small BZ, and, therefore, all symmetry transformations map
it either into itself, or into three other wavevectors $(-
\frac{\pi}{8}, \frac{\pi}{8})$, $(\frac{\pi}{8}, -\frac{\pi}{8})$ or $(-
\frac{\pi}{8}, -\frac{\pi}{8})$, all of which are equivalent in the
sense that they are connected by vectors of the reciprocal lattice.
Although Hamiltonian~(\ref{H}) is not time-reversal invariant, it is
symmetric with respect to transformation
$\hat{\tau}_{(4,4)}\mathcal{T}$, where $\mathcal{T}$ is the
time-reversal operator. Importantly, the operator
$\hat{\tau}_{(4,4)}\mathcal{T}$ transforms the wave vector ${\bf k}_0 =
(\frac{\pi}{8}, \frac{\pi}{8})$ into an equivalent wave vector
$(-\frac{\pi}{8}, -\frac{\pi}{8})$. As shown in Appendix~\ref{appendix:
4times}, this leads to the desired 4-times degeneracy.

The above proof implies that, at the wave vector $\mathbf{k}_0 = (\frac{\pi}{8}, \frac{\pi}{8})$, one double-degenerate energy band touches another double-degenerate energy band, which generally leads to a linear spectrum near the touching point, i.e. the touching energy bands have a conical shape near $\mathbf{k}_0$ -- see Fig.~\ref{fig: bands} and Fig.~\ref{fig: 3Dbands}.

We also can generalize our Hamiltonian by including additional terms, such as the ones that induce charge density modulations and/or superconductivity. Cones are robust to any such terms, provided they commute with $\hat{T}_x$, $\hat{T}_y$ and $\hat{\tau}_{(4,4)}\mathcal{T}$. One such an obvious example is a potential proportional to $|\vec{B}_{i,j}|^2$ acting on charge density.

\subsection{Plaquette-centered checkerboard}
The case $\phi_1 = \phi_2 = \frac{\pi}{8}$ in Eq.~(\ref{B}) corresponds to the plaquette-centered checkerboard shown in Fig.~\ref{fig: PLattice}. This lattice possesses unique symmetries, and, as we show in Appendix~\ref{appendix: 8times}, these symmetries result in 8-times degeneracy of each energy level at the wave vector $\mathbf{k}_0 = (\frac{\pi}{8}, \frac{\pi}{8})$.

\section{Two scenarios of pseudogap}
\label{sec: pseudogap}

We assume that, in real materials, the pseudogap in one-particle density of states $\nu(E)$ around the Fermi energy $E_F$ arises from the same energy balance that simultaneously determines the amplitude of the spin modulation. Therefore, in terms of the model description, we, in the following, first obtain $\nu(E)$ for fixed values of $t$, $t'$, $t''$ and $B_0$, then identify a dip associated with the pseudogap, and then choose the concentration of fermions such that $E_F$ corresponds to the minimum of that dip.

We consider two scenarios for the origin of the pseudogap: the ``conical-point scenario'' and ``band-edge scenario''.

\subsection{Conical-point scenario}
In Section~\ref{sec: Symmetry}, we have shown that there are cones in the electronic energy spectrum, and, therefore, one would expect that, near these conical-touch points, $\nu(E)$ is suppressed, and, thus, emergence of the pseudogap is associated with chemical potential being pinned at one of these points. For the parameters choice relevant to lanthanum cuprates, cones are not likely to be isolated in the sense that, at such a chemical potential, there are additional contributions from regular Fermi surface pockets. This would make the conical-point scenario not very different from the more general ``band-edge'' scenario described below. Yet, as discussed in Section~\ref{sec: Discussion}, the isolated conical point scenario might be realized if additional terms are included in the model Hamiltonian.

\subsection{Band-edge scenario}
In general, the checkerboard modulation does not lead to clear energy gaps between the energy bands. As the modulation amplitude $B_0$ increases, some of the energy bands develop a clear gap between themselves, while other bands still have states within that gap. This results in an incomplete suppression of the density of states, which we associate with the ``band-edge'' scenario.

Section~\ref{sec: Spin-vortex} illustrates this scenario on the basis of a concrete calculation.

\section{Calculations for a band-edge scenario}
\label{sec: Spin-vortex}

\subsection{Choice of parameters and density of states}
Here we focus on the site-centered case $\phi_1 = \phi_2 = 0$ -- see Fig.~\ref{fig: MLattice}. Below, following Ref.~\cite{pavarini2001band}, we fix $t=-1$, $t'=-0.17 t$ and $t'' = -0.5 t'$. For comparison with experiments, energy unit $1$ corresponds to approximately $350 \textmd{meV}$. Our estimation for the magnetic field amplitude is $B_0 = 0.5$.

\begin{figure}[tb!]
\centering
\def\big{\includegraphics[height=6cm]{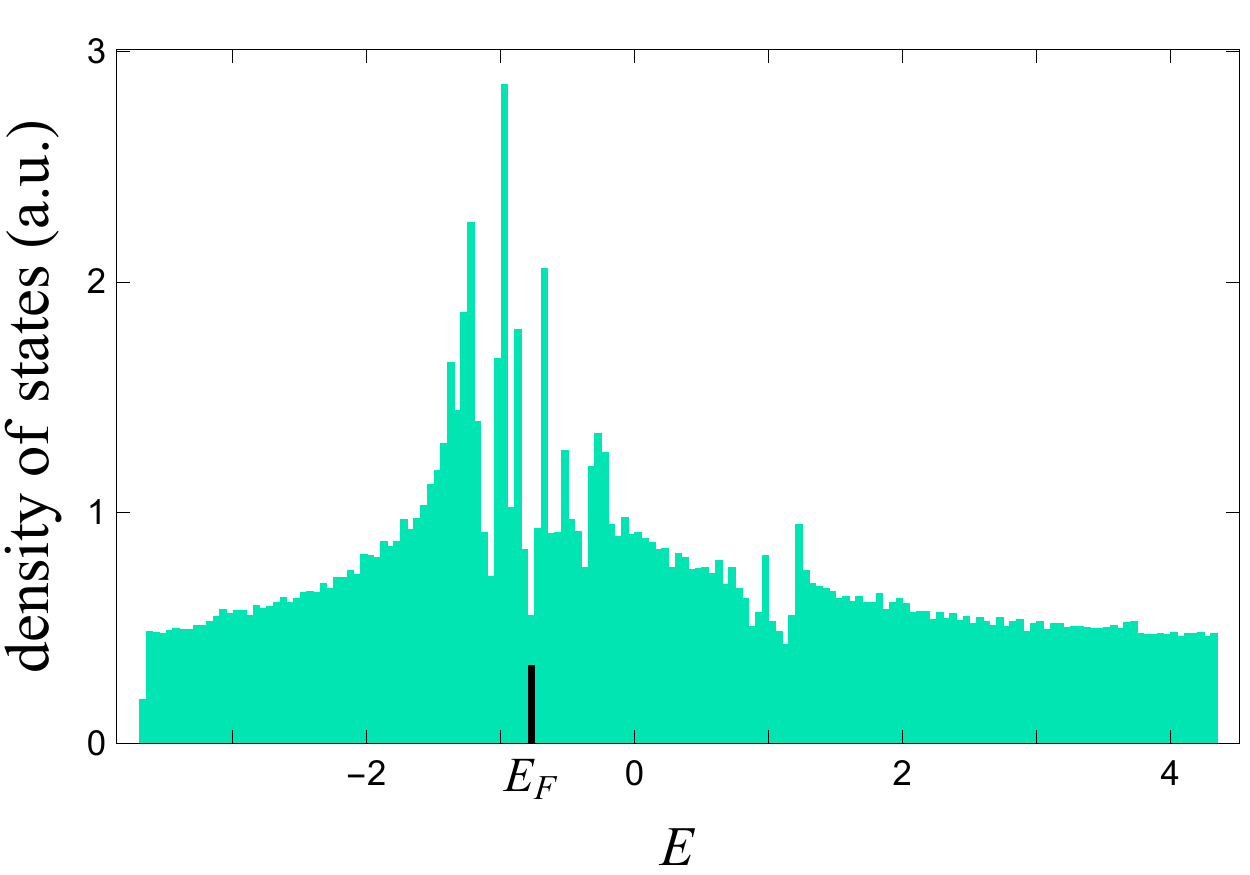}}
\def\little{\includegraphics[height=2.8cm]{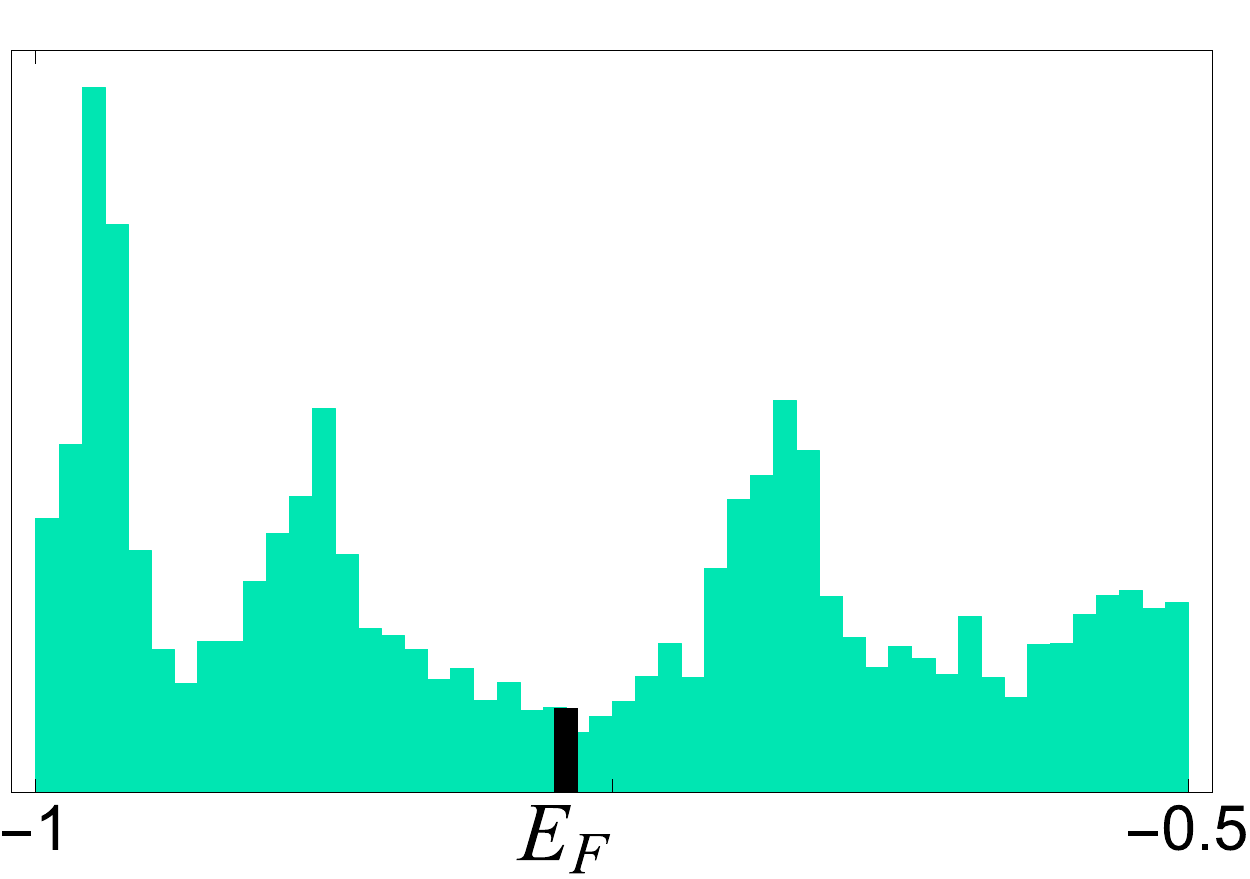}}
\def\stackalignment{r}
\topinset{\little}{\big}{6pt}{2.5pt}
\caption{(Color online). Density of states in our model for $t = -1, t' = 0.17, t'' = -0.5 t'$, and $B_0 = 0.5$. Marker indicates minimum of the DOS. Inset zooms the region close to the dip}
\label{fig:DOS}
\end{figure}

The density of states $\nu(E)$ for the above choice of parameters is shown in Fig.~\ref{fig:DOS}.

Following the approach outlined in Section~\ref{sec: pseudogap}, we place the Fermi level at $E_F = -0.77$, which, as shown in Fig.~\ref{fig:DOS}, is located at the deepest minimum in $\nu(E)$ in the energy range approximately expected to correspond to the hole-doping level 1/8. We identify this dip with the pseudogap. Such a choice leads to the concentration of fermions equal to 0.849 per site, which is reasonably close to 0.875 expected for 1/8-doped lanthanum cuprates. The discrepancy here is not of significant concern, since the concentration depends on the properties of the model far from the Fermi level, where the model is not supposed to be quantitatively accurate.  We have numerically computed the amplitude of the spin modulation, associated with the above concentration, to be equal approximately $0.3 \cdot 1/2$, which is consistent with the spin modulation amplitude reported by muon-spin relaxation ($\mu$SR) experiments~\cite{kojima2000magnetism}. This justifies our choice of the magnetic field amplitude $B_0$.
 
We further note that spin superstructure necessarily leads to charge density modulation $\delta n_{i,j}$ proportional to $|\mathbf{B}_{i,j}|^2$. The amplitude of this modulation obtained numerically is approximately $2\%$.

\begin{figure}
\includegraphics[scale=0.6]{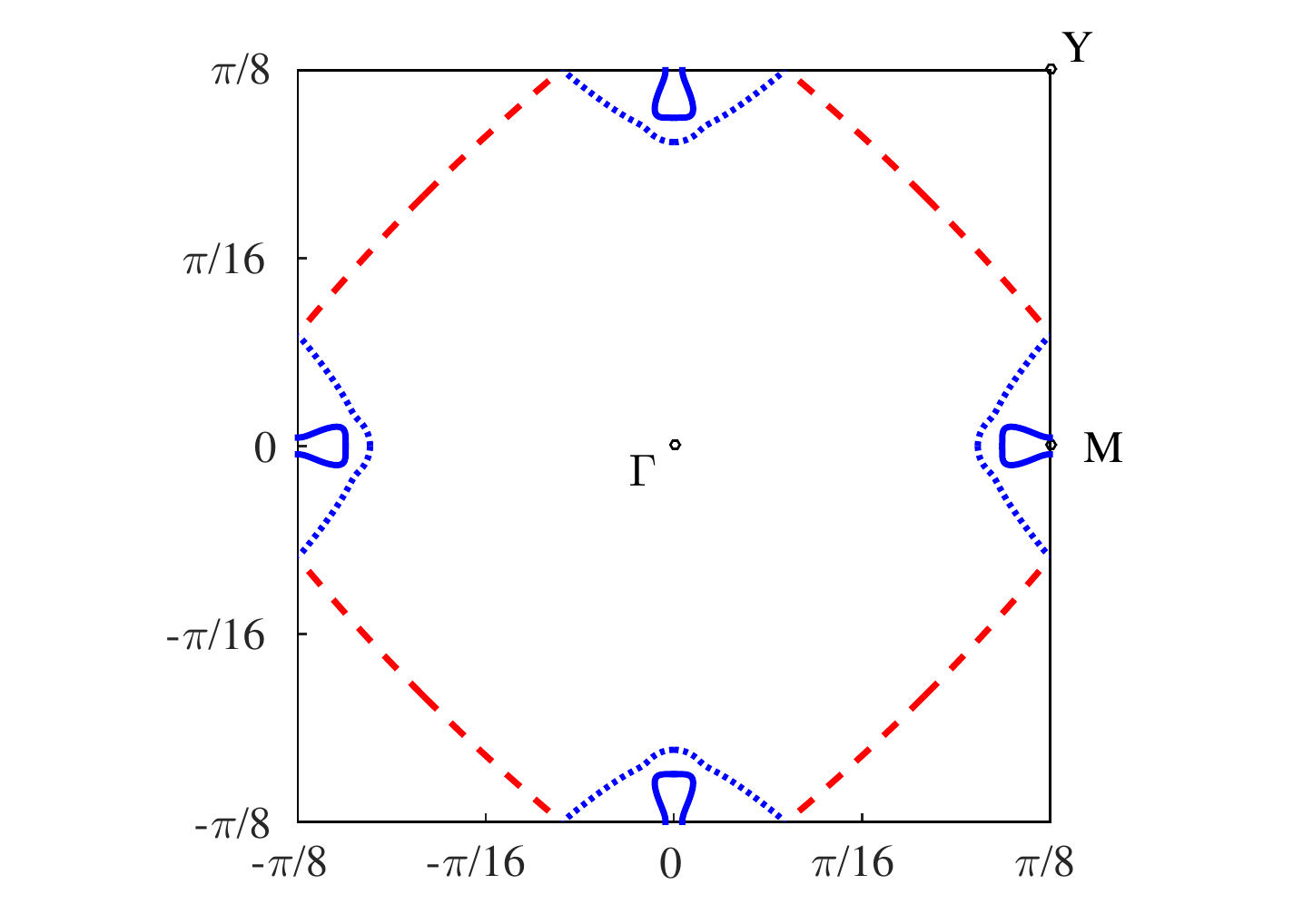}
\caption{(Color online). Calculated Fermi surface in the small BZ. It consists of 3 disjoint pockets: the largest pocket represented by the dashed red line is electron-like; the two smaller pockets represented by the dotted blue line and solid blue line are hole-like. The two larger pockets (electron-like and dotted hole-like) almost touch each, therby nearly forming a joint Fermi surface  that has the character of a connected network in the momentum space.}
\label{fig: smallFS}
\end{figure}

\begin{figure*}
\begin{minipage}[h]{0.55\linewidth}
\center{\includegraphics[width=0.85\linewidth]{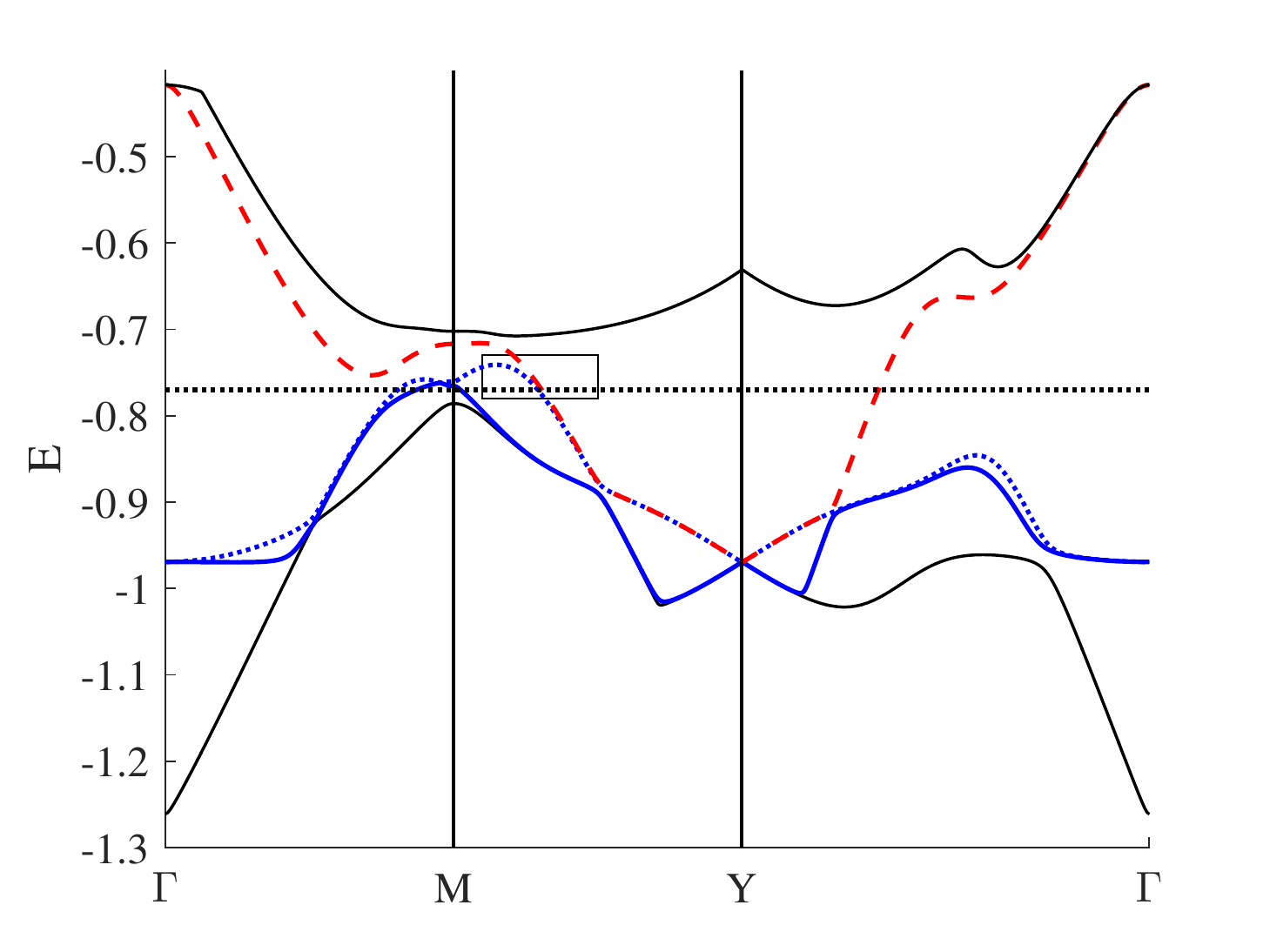}} \\a) \\
\end{minipage}
\hfill
\begin{minipage}[h]{0.4\linewidth}
\center{\includegraphics[width=1\linewidth]{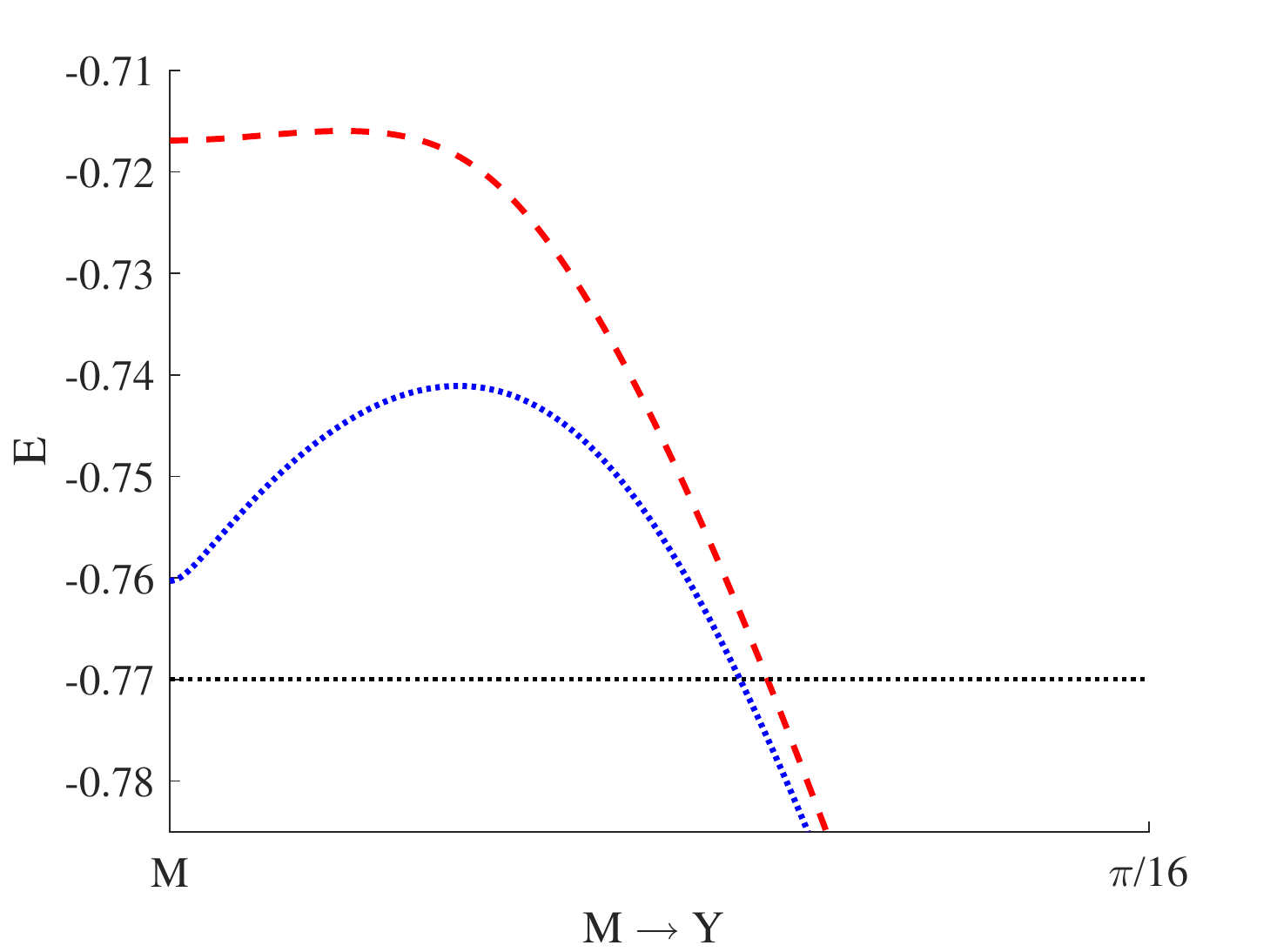}} \\b)
\end{minipage}
\caption{(Color online). a) Cuts of energy bands in the small BZ close to the Fermi level (horizontal line). Fermi surface originates from three bands; nevertheless, contribution from one of the bands is relatively small -- compare with Fig.~\ref{fig: smallFS}. All three bands originate from the same cone at the $Y$-point. b) Zoomed region in the rectangle in (a). We see that indeed the largest hole pocket and the electron pocket arise from different bands}
\label{fig: bands}
\end{figure*}

\subsection{Fermi surface in the small Brillouin zone}

\begin{figure}
\centering
\includegraphics[scale=0.5]{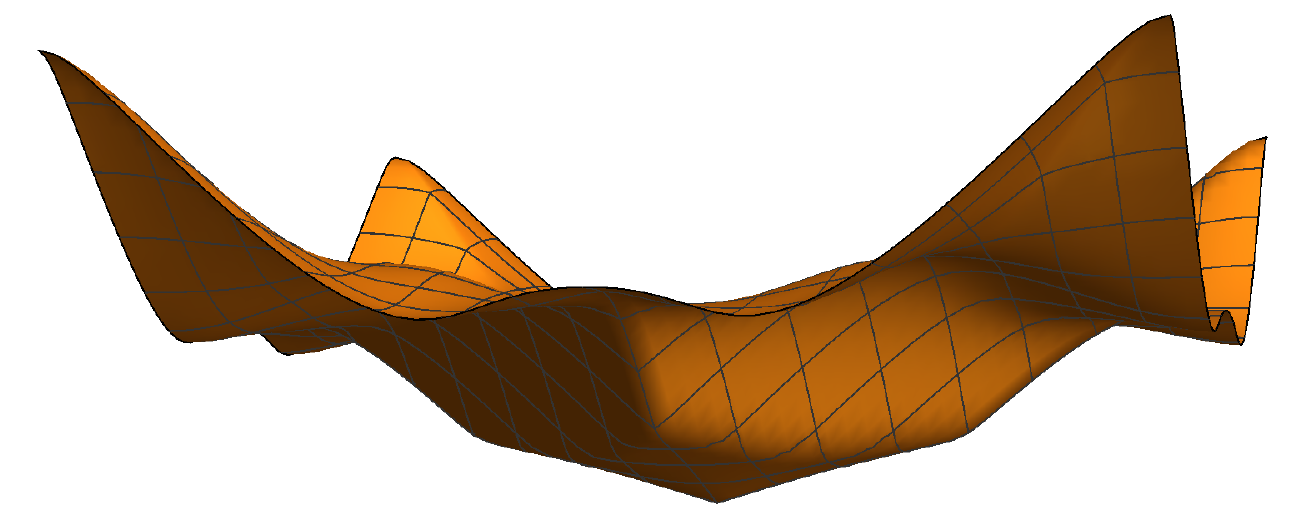}
\includegraphics[scale=0.5]{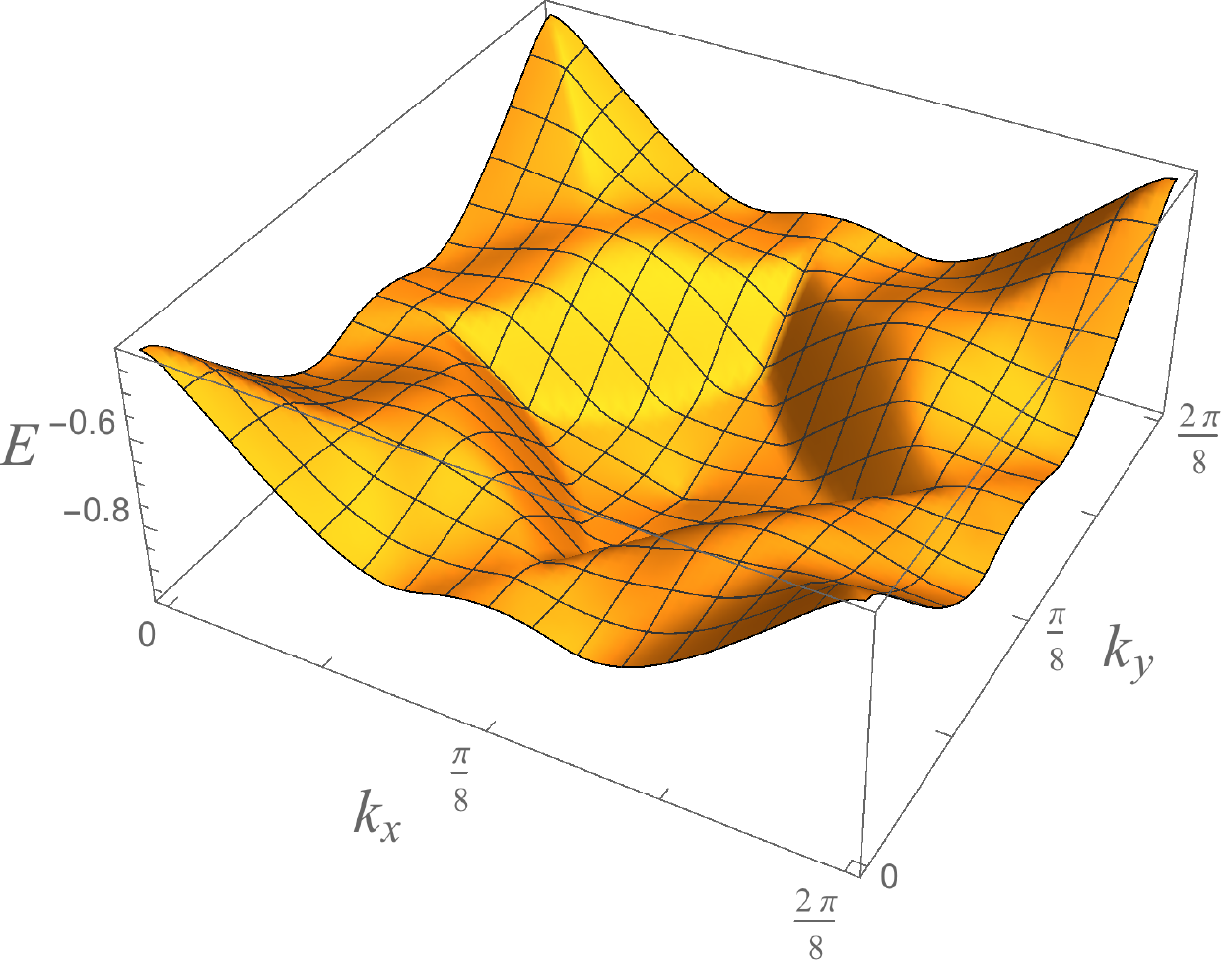}
\caption{(Color online). Different views of 3D band corresponding to the red-dashed line in Fig.~\ref{fig: smallFS} and Fig.~\ref{fig: bands}. Upper figure shows cone at $\mathbf{k}_0 = (\frac{\pi}{8}, \frac{\pi}{8})$}
\label{fig: 3Dbands}
\end{figure}

Next, we obtain the Fermi surface in the small BZ. It is shown in Fig.~\ref{fig: smallFS}. The Fermi surface consists of three disjoint parts: large electron-pocket (with area equal to $\sim 0.4 \%$ of the total area of the large BZ) and two small hole-pockets (largest hole-pocket is of size $\sim 0.05 \%$ of the area of the large BZ). Figure~\ref{fig: smallFS} may convey an incorrect impression that the two larger pockets touch each other, and, therefore, the Fermi surface forms a connected network in the $k$-space. In Fig.~\ref{fig: bands} we demonstrate that electron pocket and hole pockets are actually disjoint. They originate from different bands. Interestingly, all three bands in Fig.~\ref{fig: bands}, which contribute to the Fermi surface, originate from the same cone at the $Y$-point ($Y = (\frac{\pi}{8}, \frac{\pi}{8})$). The cone for one of the bands (dashed red line in Fig.~\ref{fig: smallFS} and in Fig.~\ref{fig: bands}) is further illustrated in Fig.~\ref{fig: 3Dbands}.

Fig.~\ref{fig: bands} shows that five bands come close to the Fermi level: three of them contribute to the Fermi surface, while the remaining two are repelled just around $E_F$. That is why we call this scenario ``band-edge.''

\subsection{Fermi surface in the large Brillouin zone}
\label{ss:largeBZ}

Let us specify the procedure of mapping the states from small BZ to the large BZ. As described in Section~\ref{sec: Model}, each eigenstate associated with a wave vector in the small BZ is represented by a superposition of initial states (eigenstates of the tight-binding Hamiltonian $\mathcal{H}_0$) that correspond to wave vectors in the large BZ:
\begin{equation}
\Ket{\tilde{\mathbf{q}}} = \sum\limits_{l,m = 0,...,7; \alpha = \pm} C_{lm, \alpha} \Ket{\tilde{\mathbf{q}} + \frac{\pi}{4}\cdot l \cdot \mathbf{e}_x + \frac{\pi}{4} \cdot m \cdot \mathbf{e}_y, \alpha}.
\label{qtilde}
\end{equation}
This gives the mapping from the small BZ to the large one: the sum $|C_{lm,+}|^2 + |C_{lm,-}|^2$ represents the spectral weight at the wave vector
$\tilde{\mathbf{q}} + \frac{\pi}{4}\cdot l \cdot \mathbf{e}_x + \frac{\pi}{4} \cdot m \cdot \mathbf{e}_y$
 in the large BZ. 
 
In order to obtain Fermi surface in the large BZ, we have chosen sufficiently fine grid of wave vectors $\tilde{\mathbf{q}}$ in the small BZ. For each $\tilde{\mathbf{q}}$, we numerically computed the eigenstates in Eq.~(\ref{qtilde}), then selected those of them that fell in the energy window  $E = -0.77\pm 0.03$ and then, for each, plotted the spectral weights of the participating wave vectors  $\tilde{\mathbf{q}} + \frac{\pi}{4}\cdot l \cdot \mathbf{e}_x + \frac{\pi}{4} \cdot m \cdot \mathbf{e}_y$.
 The result of such a mapping is shown in Fig.~\ref{fig: largeFS}.  In Appendix~\ref{appendix: rec}, we further illustrate contributions in the large BZ from each of the three Fermi surface pockets present in the small BZ.
 
\begin{figure}
\begin{overpic}[width=0.8\linewidth,tics=10]{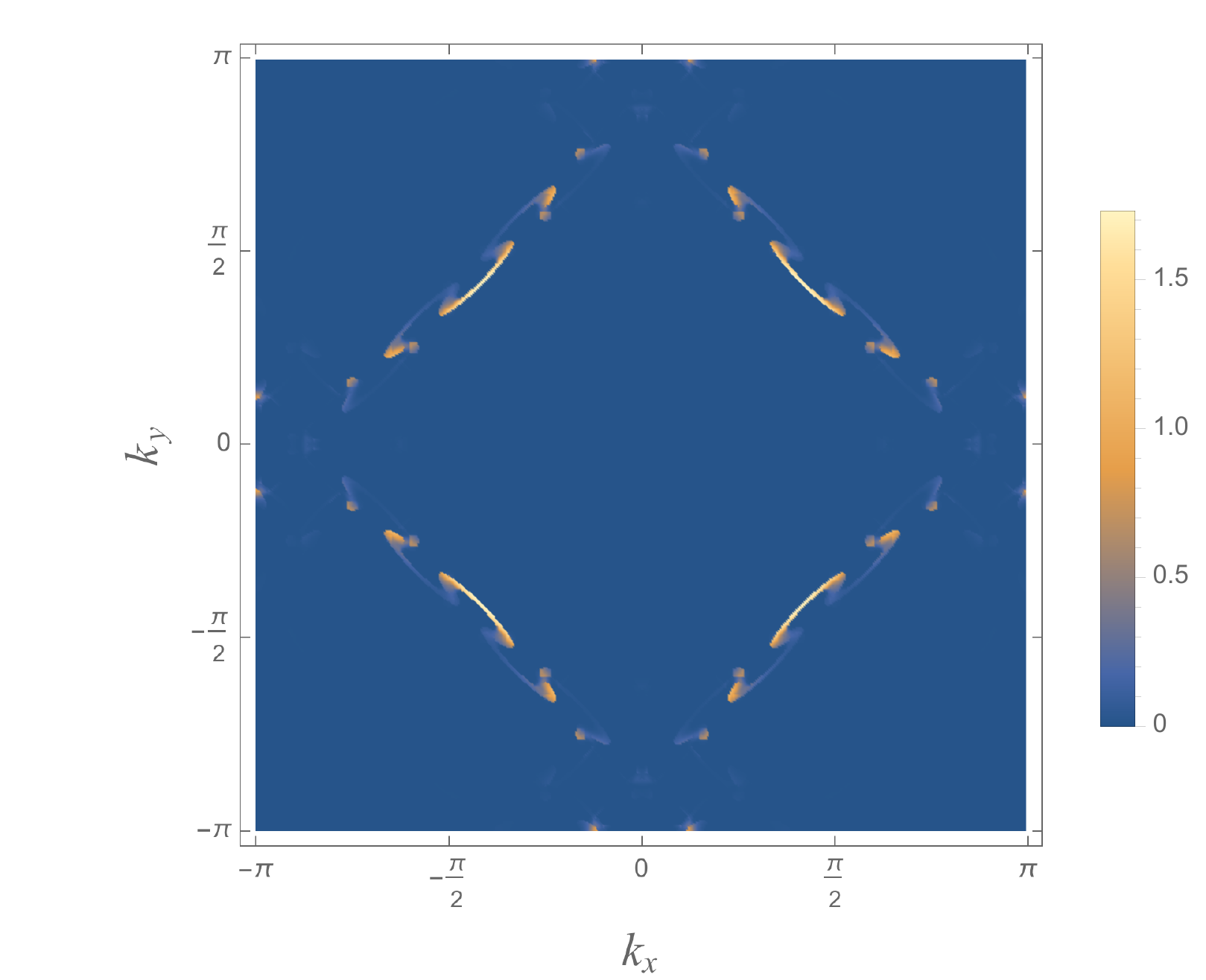}
 \put (87,68){ \rotatebox{0}{[a.u.]}}
\label{fig: largeFS}
\end{overpic}
\caption{(Color online). Reconstructed Fermi surface in the large BZ. The procedure of its reconstruction is specified in the text}
\label{fig: largeFS}
\end{figure}
 
In Fig. ~\ref{fig: largeFS}, one can clearly see Fermi arcs in the nodal directions.  Other spots of lower intensity also appear, but, so far, they have not been observed in experiments. The comparison of Figs.~\ref{fig: largeFS} and~\ref{fig: pockets} indicates that the Fermi arcs originate from the largest Fermi surface pocket in the small BZ, while the remaining spots originate from the two smaller pockets.  These smaller pockets are likely related to the non-interacting character of our model. They may possibly be removed if superconducting fluctuations are introduced -- see Ref.~\cite{allais2014connecting}.

\section{Discussion}
\label{sec: Discussion}

Let us first consider the isolated conical-point scenario described in Section~\ref{sec: pseudogap}. It requires either a large value of the local field amplitude $B_0$ or the inclusion of extra terms in the Hamiltonian not considered in the present article, such as those associated with charge-density and lattice modulations~\cite{egami2010spin} and/or superconducting fluctuations~\cite{allais2014connecting}, provided these terms would respect the symmetries discussed in Section~\ref{sec: Symmetry}. They may further separate energy bands and, as a result, isolate cones. In real materials, contributions from such terms might be large, and, hence, the isolated conical-point scenario would become relevant.

If this scenario is realized then it would lead to the absence of quantum oscillations, because in such a case the Fermi surface in the small BZ would be reduced to a single point.

The band-edge scenario leads to parameter-dependent predictions. We can now analyze the example computed in Section~\ref{sec: Spin-vortex} and then draw general lessons from it.

\begin{figure}
\hspace*{-0.98cm}\begin{minipage}[h]{0.45\linewidth}
\includegraphics[width=1.6\linewidth]{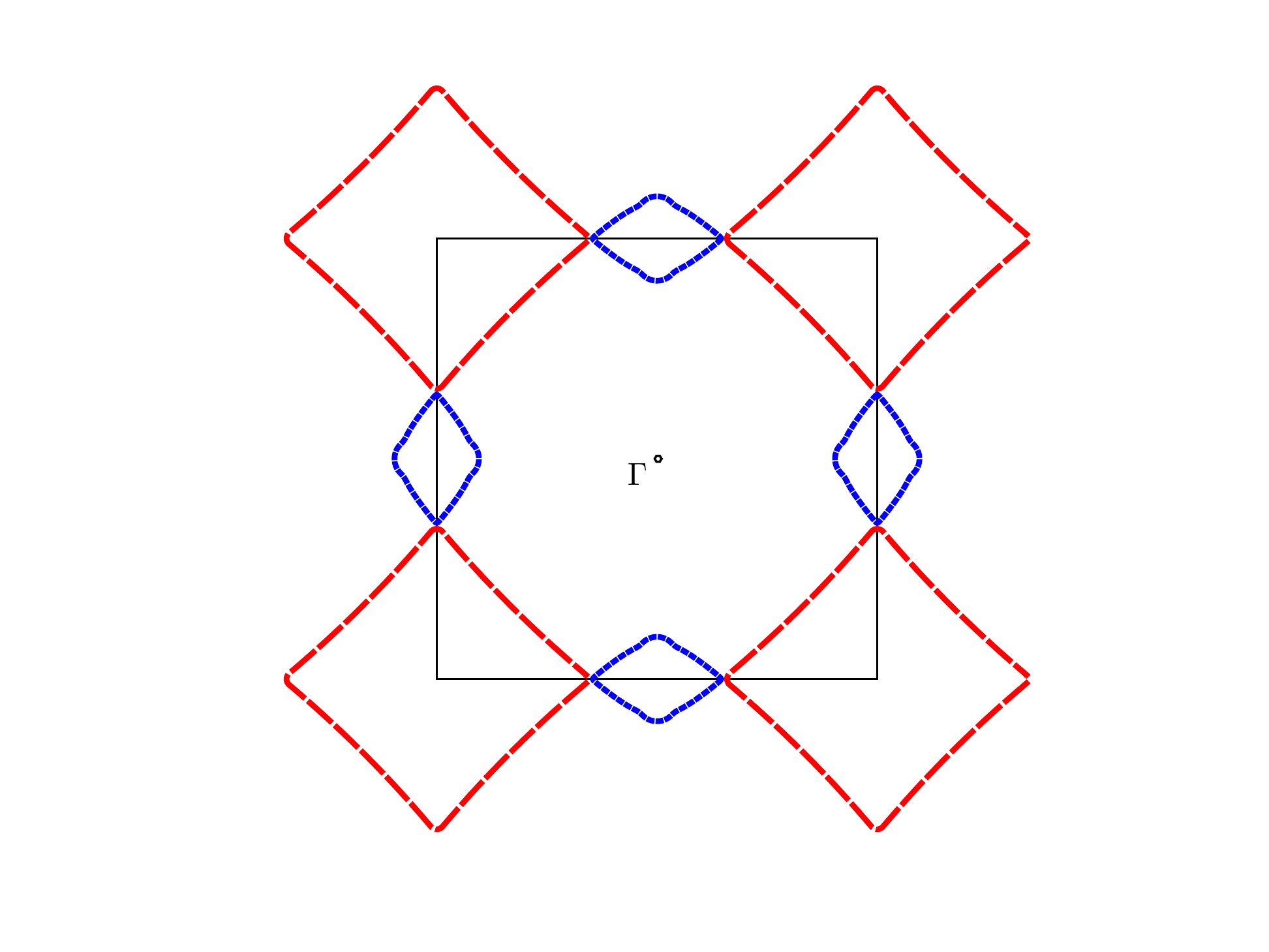} \\a) \\
\end{minipage}
\hspace*{0.95cm}
\begin{minipage}[h]{0.45\linewidth}
\includegraphics[width=1.6\linewidth]{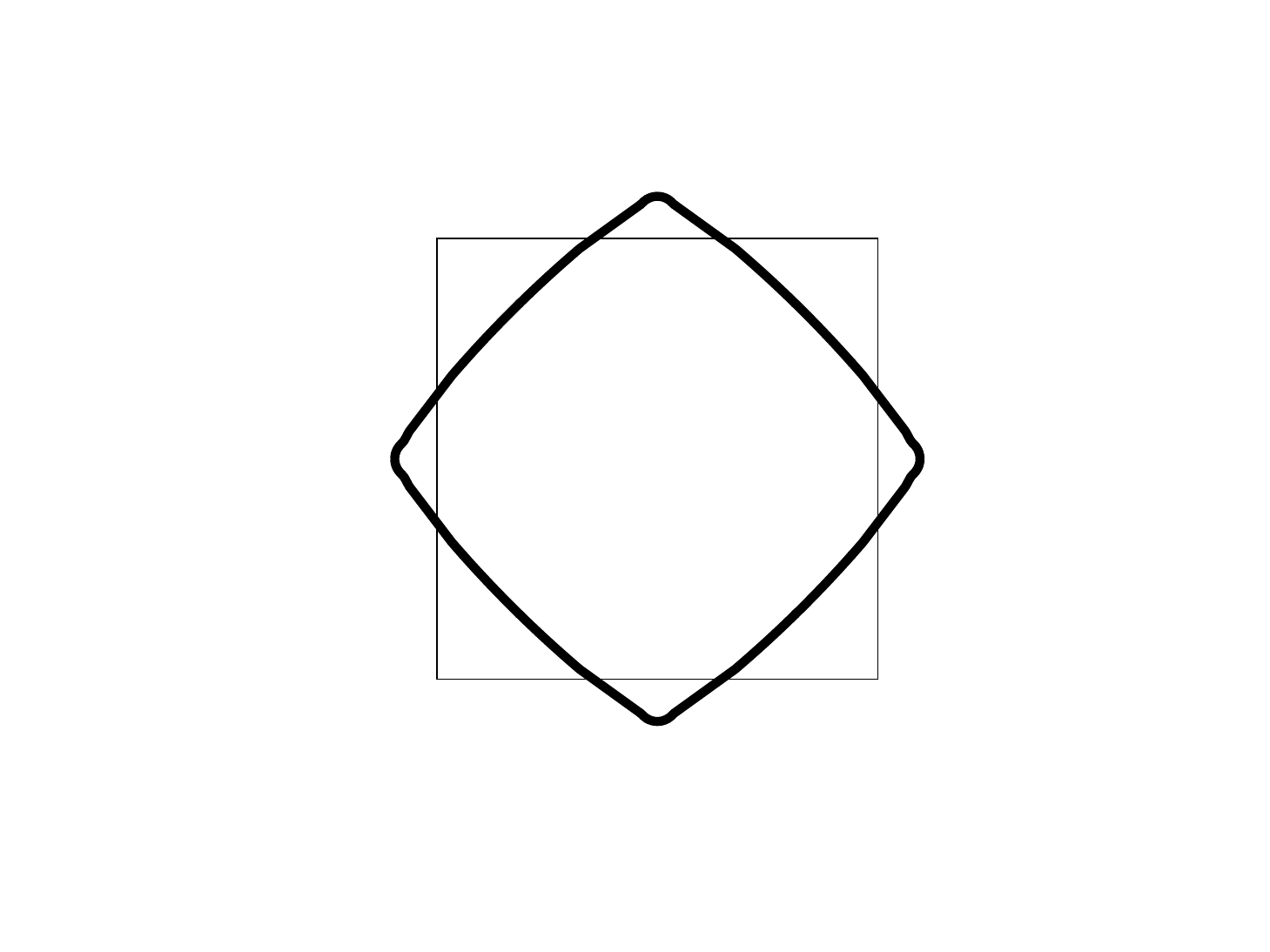} \\b)
\end{minipage}
\caption{(Color online). a) Quantum oscillations at low magnetic field $H\ll H_c$ contain contributions from both electron pocket depicted by dashed red line and the larger hole pocket depicted by dotted blue line (compare with Fig.~\ref{fig: smallFS}); b) At large magnetic field $H\gg H_c$, semi-classical wave packets on the Fermi surface move as if the two pockets in (a) merge into a single Fermi surface depicted by solid black line.}
\label{fig: magneticBreakdown}
\end{figure}

Fermi surface in Fig.~\ref{fig: smallFS} contains two pockets, which, at the BZ boundary, almost touch each other. (Here, we neglect the smallest Fermi surface pocket.) This suggests the possibility of magnetic breakdown between the corresponding bands. We estimate the characteristic field $H_c$ for the onset of magnetic breakdown from the condition~\cite{ashcroft1976solid}: $\frac{e\hbar H_c}{m c} \sim \frac{\epsilon^2_g}{W}$, where $W \sim 0.3  t$ is characteristic bandwidth (see Fig.~\ref{fig: bands} (a)), and $\epsilon_g$ is a gap at $E_F$ between the two bands. As one can see in Fig.~\ref{fig: bands} (b), $\epsilon_g\lesssim 0.03 \cdot t$, resulting in $H_c \lesssim 10\textmd{T}$. Fig.~\ref{fig: magneticBreakdown}(a) illustrates electron and hole pockets contributing to quantum oscillations at low magnetic field $H \ll H_c$. Due to smallness of each pockets, observations of quantum oscillations would probably require samples with unrealistically long quasi-particle lifetime. In the opposite limit $H \gg H_c$, semi-classical fermionic wave packets will follow the trajectory in the momentum space shown in Fig.~\ref{fig: magneticBreakdown}(b). This trajectory switches between the available Fermi surface pockets. This signal is hole-like with the total area of $\sim 1.3 \%$ of the large BZ, large enough to be detectable. 

 We can now draw general lessons from the above example. The spin-vortex checkerboard modulation in the interesting range of parameters produces quite a dense set of energy bands with many symmetries. It is therefore to be expected that these bands have quite a few avoided crossings, which, in turn, if sliced at constant energy, would lead to multiple Fermi surfaces nearly touching each other. Such a pattern of Fermi surfaces is likely to suppress quantum oscillations, because of multiple points, where at moderate external magnetic fields, magnetic breakdowns may occur, so that effectively the Fermi surface is turned into an open network in the momentum space without well-defined cyclotron frequency.

In the both of the above scenarios, one possible way to explain the drop of resistivity reported in Refs.~\cite{li2007two,tranquada2008evidence} is to attribute it to a first-order-like crystallization of spin superstructure, which suddenly increases mean-free path of quasi-particle excitations. This would be similar to what occurs with simple metals as they undergo a first-order crystallization transition. In the framework of the conical-point scenario, the possibility of the resistivity drop is further strengthened by the fact that, like in graphene, a Fermi surface reduced to a few conical points suppresses the scattering of the quasi-particles and hence should significantly increase their mobility. The alternative interpretation is the one proposed by the authors of Ref.~\cite{li2007two} and supported by the strong magnetic-field dependence of the resistivity drop, namely that it is caused by the onset of two-dimensional fluctuating superconductivity. This interpretation as such does not disciminate between stripes and checkerboards. Superconductivity in the presence of stripes was considered in Ref.~\cite{Berg2007},  while, for the spin-vortex checkerboard, it was done in Ref.~\cite{Bhartiya2017}.

The multitude of the bands arising for the spin-vortex checkerboard, and their dependence on the model parameters, prevent us from making definite predictions about the Seebeck coefficient. If, however, our assumption that the minimization of the total energy of the system requires the modulation parameters to adjust themselves in such a way that the chemical potential becomes pinned at the bottom of the pseudogap, then this implies that the density of states is close to being symmetric on the both sides of the chemical potential, which in turn, would suggest that the Seebeck coefficient is close to zero and can easily change sign as a function of temperature or doping. This is indeed what is observed experimentally -- see Ref.~\cite{laliberte2011fermi}.  

Finally, we would like to remark, that one of likely features of spin-vortex checkerboard modulations irrespective of a particular scenario is that {\it more than one band} come close to the Fermi surface. It is, therefore, to be expected that such features are to be seen by ARPES. This proposition is consistent with the ARPES experiment of Ref.~\cite{chang2008electronic} reporting the observation of two bands for a particular momentum cut through the BZ.

\section{Conclusions}
\label{sec: Conclusions}

We have calculated band structure for the model of non-interacting fermions in the background of spin-vortex checkerboard and analyzed symmetry properties and degeneracies of the resulting bands. We have proven that each band is double degenerate and in addition  has at least one conical point where it touches another double-degenerate band. We then considered two scenarios for the emergence of the pseudogap: (i) the conical-point scenario and (ii) the band-edge scenario. For the model parameters estimated to be relevant to $1/8$-doped lanthanum cuprates, the isolated conical-point scenario is not realizable, because the Fermi surfaces corresponding to energies of each of the available conical points also contain additional regular pockets. The conical feature is, nevertheless, robust, because it is symmetry-protected. Therefore, the conical-point scenario may become relevant if the model Hamiltonian is further generalized to include terms representing charge modulations and superconducting fluctuations. As for the band-edge scenario, we performed a concrete calculation, which led to the Fermi surface containing Fermi arcs along the nodal directions -- in agreement with experiments -- and in addition, some low-intensity spots not observed experimentally. Our analysis indicates that quantum oscillations of transport coefficients would be suppressed in the presence of spin-vortex checkerboard within either of the above two scenarios. It also appears that our model is largely consistent with the measurements of resistivity and Seebeck coefficient in $1/8$-doped lanthanum cuprates. 

\section{Acknowledgements}
We would like to thank G. A. Starkov and A. A. Katanin for discussions. This project was funded by Skoltech as a part of Skoltech NGP program.


\appendix

\section{4-times degeneracy at ${\bf k}_0 = (\frac{\pi}{8},\frac{\pi}{8})$}
\label{appendix: 4times}

Let us consider one fermion on the spin-vortex checkerboard lattice and parametrize its wave function as follows:
\begin{equation}
\psi(x,y) = u(x,y) \begin{pmatrix}
1 \\
0
\end{pmatrix}
+v(x,y) \begin{pmatrix}
0 \\
1
\end{pmatrix},
\label{eqn: wavefunction}
\end{equation}
where $u(x,y)$ and $v(x,y)$ are spatial functions defined on the two-dimensional lattice plane; $\begin{pmatrix}
1 \\
0
\end{pmatrix}$ and $\begin{pmatrix}
0 \\
1
\end{pmatrix}$ correspond to spin projections on the $z$-axis.

For convenience, we also introduce operator
\begin{equation}
\hat{T}_{xy} \equiv i \hat{T}_x \hat{T}_y = \hat{\tau}_{(4,4)} \otimes
\sigma_z
\label{Txy}
\end{equation}
representing  translation by vector $(4,4)$ with subsequent rotation of
spins through $180^{\circ}$ about the $z$-axis.

Let us recall that the time-reversal operator $\mathcal{T}$ acts on a wave function given by Eq.~(\ref{eqn: wavefunction}) as follows:
\begin{equation}
\mathcal{T} \psi = i u^{*}(x,y) \begin{pmatrix}
0 \\
1
\end{pmatrix}
-i v^{*}(x,y) \begin{pmatrix}
1 \\
0
\end{pmatrix}.
\label{eqn: T_R}
\end{equation}
Note that $\mathcal{T}^2 = -1$. The important property of the time-reversal operator is that it reverses both the spin and the wave vector.

One can check the anti-commutation relation $\{ \hat{\tau}_{x,y} \mathcal{T}, \hat{T}_{x,y} \} = 0$. 

From now on, we focus our attention on functions $u(x,y)$ and $v(x,y)$ that correspond to $\mathbf{k}_0 =(\frac{\pi}{8},\frac{\pi}{8})$. From Bloch's theorem, it follows that such functions are anti-periodic with respect to translation by 8 lattice constants, i.e. $u(x+8,y) = u(x,y+8) = -u(x,y)$. Note that, for such wave functions, $\hat{T}_x^2 = \hat{T}_y^2 =  \hat{T}^2_{x,y} = 1$. 
By analogy to spin operators, we introduce operators $\hat{Q} = \hat{T}_x + i \hat{T}_y$, $\hat{Q}^{\dagger} =  \hat{T}_x - i \hat{T}_y$, which can be considered as raising and lowering operators while acting on eigenstates of operator $\hat{T}_{x,y}$:
\begin{equation}
\Big[ \hat{T}_{x,y}, \hat{Q}^{(\dagger)} \Big] = \mp 2 \hat{Q}^{(\dagger)}. 
\label{eqn: commut}
\end{equation}
Since each energy level is double degenerate, it is convenient to characterize each eigenstate by two quantum numbers -- energy and eigenvalue $\lambda$ of the operator $\hat{T}_{x,y}$, which can take values $\pm 1$.

Let us consider an energy eigenstate $\psi$ with $\lambda = 1$, i.e. $\hat{T}_{x,y} \psi = \psi$. From $\psi$ we can construct new state $\tilde{\psi} = \hat{\tau}_{x,y}\mathcal{T} \circ \hat{Q} \psi$, which has the same energy and the same $\lambda = 1$. The fact that $\tilde{\psi}$ has the same $\lambda = 1$ follows from the fact that the operator $\hat{Q}$ lowers $\lambda$ to become $-1$; but, since operators  $\hat{\tau}_{x,y}\mathcal{T}$ and $\hat{T}_{x,y}$ anti-commute, $\hat{\tau}_{x,y}\mathcal{T}$ raises $\lambda$ back to be 1. Our goal now is to show that $\psi$ and $\tilde{\psi}$ are two linearly independent states. This, together with our previous argument for 2-times degeneracy of each energy band, will prove the desired 4-times degeneracy.

Using the definition of the operator $\hat{Q}$, one can check that
\begin{eqnarray}
\psi_1 &=& \hat{Q} \psi = i (\hat{\tau}_{4x}u - \hat{\tau}_{4y}u) \begin{pmatrix}
0\\
1
\end{pmatrix} + i (\hat{\tau}_{4x}v + \hat{\tau}_{4y}v) \begin{pmatrix}
1\\
0
\end{pmatrix} = \notag{} \\
&=& 2i\hat{\tau}_{4x}u \begin{pmatrix}
0\\
1
\end{pmatrix} + 2i \hat{\tau}_{4x}v \begin{pmatrix}
1\\
0
\end{pmatrix},
\end{eqnarray}
where, in the last equality, we implied that $\hat{\tau}_{x,y} u = u$ and $\hat{\tau}_{x,y} v = -v$, which follows from $\hat{T}_{x,y} \psi = \psi$. Using Eq.~(\ref{eqn: T_R}), we then obtain:
\begin{eqnarray}
\tilde{\psi} = \hat{\tau}_{x,y}\mathcal{T} \psi_1 = -2 (\hat{\tau}_{4x}u)^* \begin{pmatrix}
1\\
0
\end{pmatrix} - 2(\hat{\tau}_{4x}v)^* \begin{pmatrix}
0\\
1
\end{pmatrix}.
\end{eqnarray}

Therefore, $\psi$ and $\tilde{\psi}$ are linearly dependent if and only if: 
\begin{eqnarray}
&& u^{*}  =  \alpha \hat{\tau}_{4x} u\\
&& v^{*} =  \alpha \hat{\tau}_{4x} v,
\end{eqnarray}
where $\alpha$ is some nonzero complex number. We can use the last two equations to obtain:
\begin{eqnarray}
 u(x,y) &=& (u^{*})^{*} = ( \alpha u(x+4, y))^{*} = 
  \alpha^{*} u^{*}(x+4,y) =\notag{}\\ 
  &=&|\alpha|^2 u(x+8, y) = - |\alpha|^2 u(x,y).
\end{eqnarray}
The same relation holds for $v$. For nonzero $u(x,y)$ and $v(x,y)$, the above equality can only be satisfied, when $\alpha = 0$, but this means that $\psi$ and $\tilde{\psi}$ are linearly independent. This completes the proof of the 4-fold degeneracy of each energy level at $\mathbf{k}_0 = (\frac{\pi}{8},\frac{\pi}{8})$.

\section{8-fold degeneracy at ${\bf k}_0 = (\frac{\pi}{8},\frac{\pi}{8})$ for the plaquette-centered checkerboard}
\label{appendix: 8times}
\begin{figure}[ht!]
\centering
\includegraphics[scale=0.3]{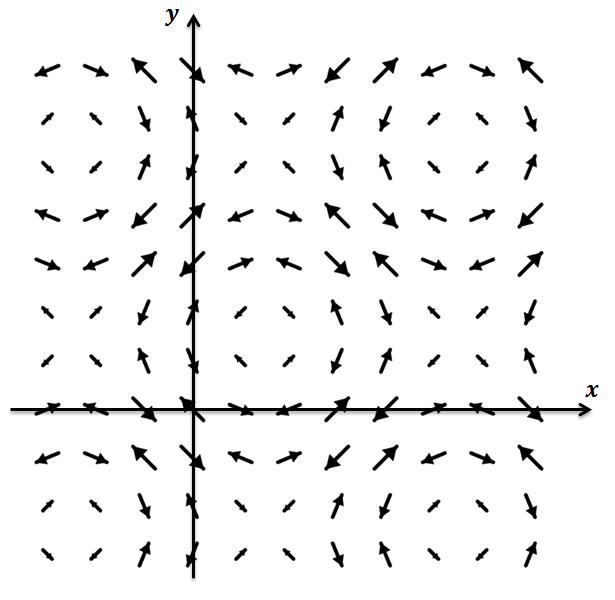}
\caption{Plaquette-centered checkerboard corresponding to $\phi_1 = \phi_2 = \pi/8$ in Eq.~(\ref{B})}
\label{fig: PLattice}
\end{figure}

Let us introduce three more symmetry transformations specific to the case of plaquette-centered checkerboard:
\begin{eqnarray}
&& \hat{A}_x \equiv \mathit{R}_x \hat{\tau}_{(0,1)} \otimes \sigma_z\\
&& \hat{A}_y \equiv \mathit{R}_y \hat{\tau}_{(1,0)} \otimes \sigma_z\\
&& \hat{S} \equiv \hat{A}_x \hat{A}_y = \hat{I}\hat{\tau}_{(1,1)},
\label{eqn: S}
\end{eqnarray}
where $\mathit{R}_{x}$ ($\mathit{R}_{y}$) denotes spatial reflection with respect to the $x$-axis ($y$-axis) shown in Fig.~\ref{fig: PLattice}; $\hat{I}$ denotes spatial inversion with respect to the coordinate origin shown in Fig.~\ref{fig: PLattice}. All three operators $\hat{A}_x$, $\hat{A}_y$ and $\hat{S}$ commute with each other and with the Hamiltonian $\mathcal{H}$.

\begin{figure*}
\begin{minipage}[h]{0.3\linewidth}
\center{\begin{overpic}[width=1.1\linewidth,tics=10]{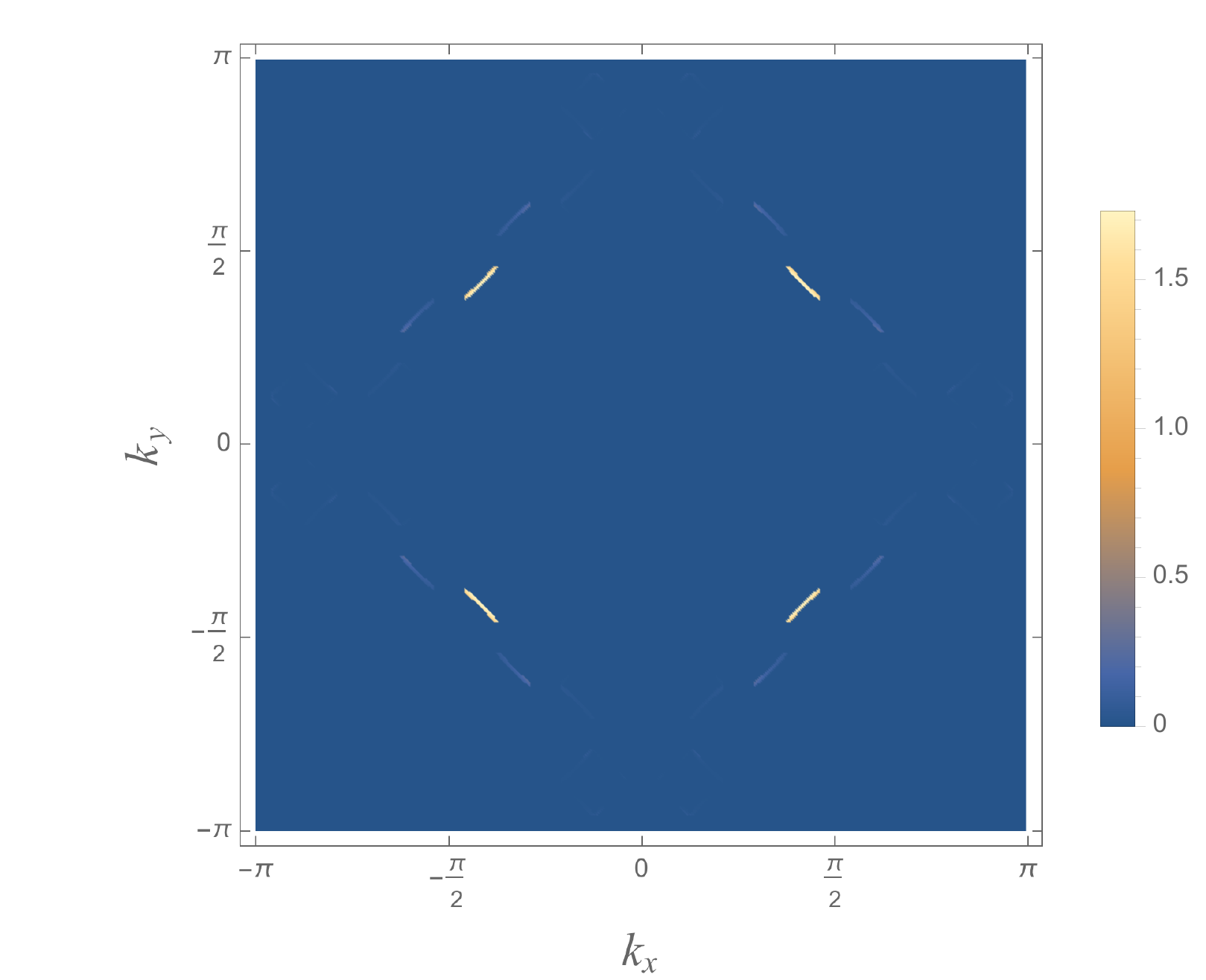}
 \put (87,68){ \rotatebox{0}{[a.u.]}}
\end{overpic}} \\a) \\
\end{minipage}
\hfill
\begin{minipage}[h]{0.3\linewidth}
\center{\begin{overpic}[width=1.1\linewidth,tics=10]{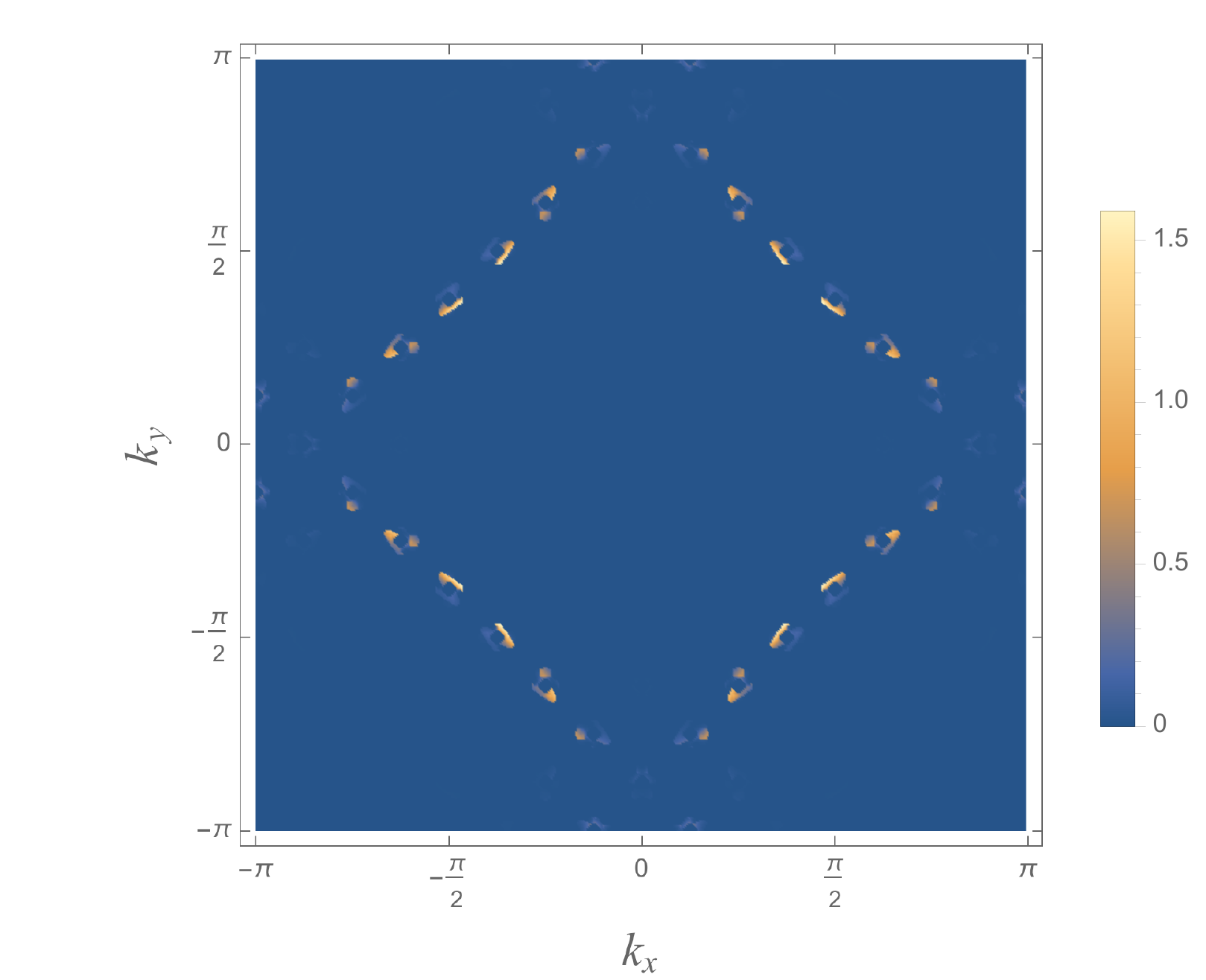}
 \put (87,68){ \rotatebox{0}{[a.u.]}}
\end{overpic}} \\b)
\end{minipage}
\hfill
\begin{minipage}[h]{0.3\linewidth}
\center{\begin{overpic}[width=1.1 \linewidth,tics=10]{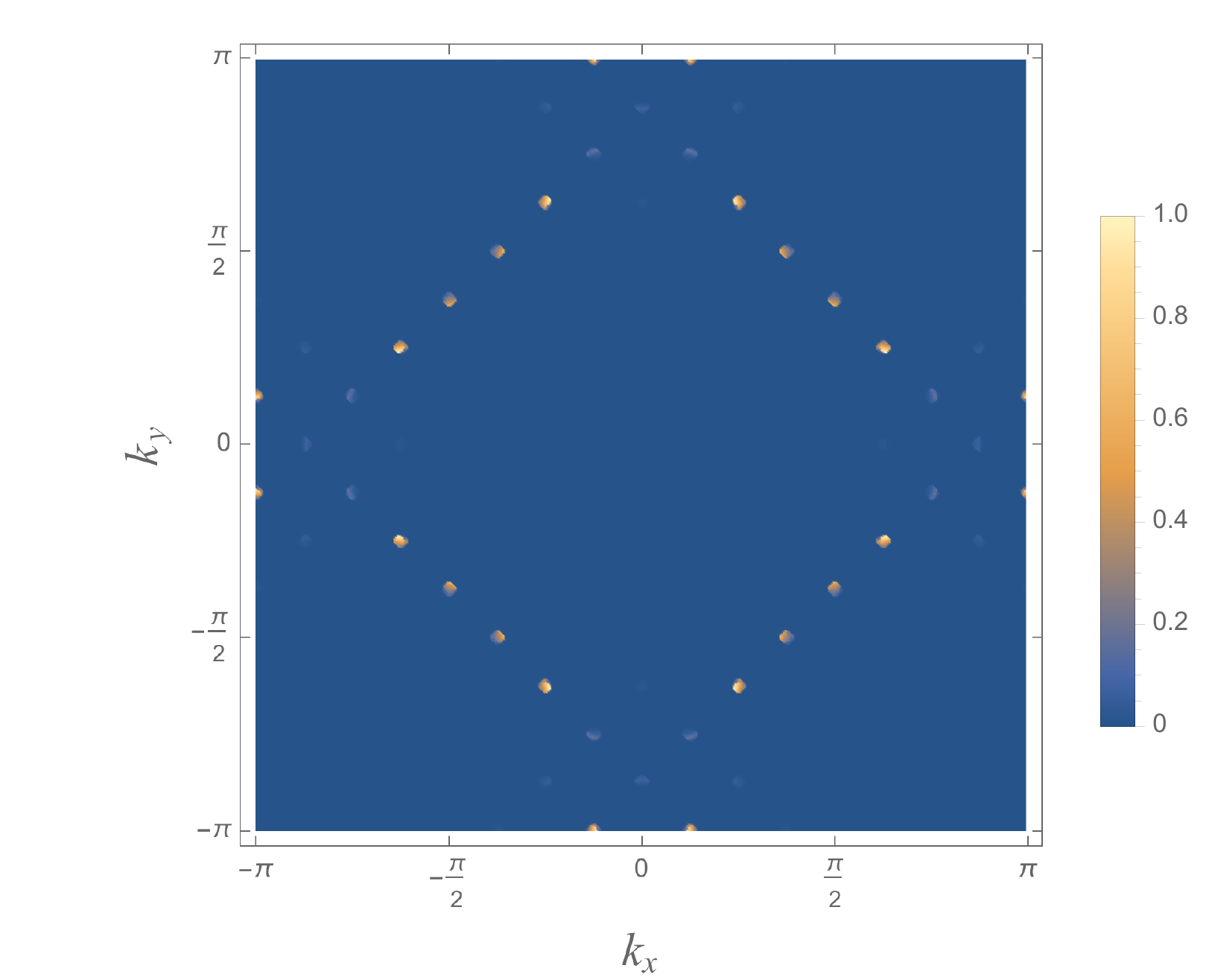}
 \put (87,68){ \rotatebox{0}{[a.u.]}}
\end{overpic}} \\c)
\end{minipage}
\caption{(Color online). Mapping of the Fermi-surface pockets from the small BZ shown in Fig.~\ref{fig: smallFS} to the large BZ:  a) the largest  pocket (red dashed line in Fig.~\ref{fig: smallFS}); b) smaller pocket (blue dashed line in Fig.~\ref{fig: smallFS}); and c) the smallest pocket (blue solid line in Fig.~\ref{fig: smallFS}).}
\label{fig: pockets}
\end{figure*}

Operators $\hat{A}_{x}$ and $\hat{A}_{y}$ anti-commute with $\hat{T}_{x,y}$. In order to show this, let us consider some function $f(x,y)$ that has spatial periodicity corresponding to $\mathbf{k}_0 = (\frac{\pi}{8},\frac{\pi}{8})$. In this case, we find:
\begin{eqnarray*}
 \mathit{R}_x \hat{\tau}_{(0,1)} \hat{\tau}_{(4,4)} f(x, y) &=& f(x+4, -y-5)\\
 \hat{\tau}_{(4,4)} \mathit{R}_x \hat{\tau}_{(0,1)}  f(x, y) &=& f(x+4, -y+3) =\notag{}\\ 
 &=&- f(x+4, -y-5),
\end{eqnarray*}
which implies that $\hat{S}$ commutes with $\hat{T}_{x,y}$, because the operator $\hat{S}$ is a product of operators $\hat{A}_x$ and $\hat{A}_y$, each of which anti-commutes with $\hat{T}_{x,y}$. As a result, each eigenstate can be characterized, in addition to $\lambda$, by a quantum number $\lambda_s$ associated with the operator $\hat{S}$. 

Consider state $\psi$ with $\lambda = 1, \lambda_s = 1$. We now observe that $\tilde{\psi} = \hat{\tau}_{x,y}\mathcal{T}\hat{A}_x \psi$ is a state with the same energy and with $\lambda = 1, \lambda_s = 1$. Indeed, both operators $\hat{\tau}_{x,y}\mathcal{T}$ and $\hat{A}_x$ anti-commute with $\hat{T}_{x,y}$, so that their product commutes with $\hat{T}_{x,y}$; both of them as well as their product commutes with $\hat{S}$. 

Let us write $\tilde{\psi}$ explicitly
\begin{align}
&\tilde{\psi} = i u^{*}(x,-y-1)\begin{pmatrix}
0\\
1
\end{pmatrix} - i v^{*}(x, -y-1)\begin{pmatrix}
1\\
0
\end{pmatrix}
\end{align}
and then prove that $\tilde{\psi}$ and $\psi$ (given by Eq.~(\ref{eqn: wavefunction})) are linearly independent. They are linearly dependent if and only if:
\begin{eqnarray}
&& u(x,y) = - \alpha v^{*} (x,-y-1)\\
&& v(x,y) = \alpha u^{*} (x, -y -1),
\end{eqnarray}
for some nonzero complex number $\alpha$. We can use the last two equations to obtain:
\begin{equation}
u(x,y) = - |\alpha|^2 u(x,y).
\end{equation}
The same identity holds for $v(x,y)$. For nonzero $u(x,y)$ and $v(x,y)$, the last identity can be satisfied only when $\alpha = 0$, but this means linear independence of $\psi$ and $\tilde{\psi}$. From this, it follows that point $\mathbf{k}_0 = (\frac{\pi}{8},\frac{\pi}{8})$ for the plaquette-centered case has an additional two-fold degeneracy, which, together with the previously proven general four-fold degeneracy, implies the overall 8-times degeneracy.

\section{Mapping of individual Fermi-surface pockets to the large BZ}
\label{appendix: rec}

In Fig.~\ref{fig: pockets}, we present the individual mapping of each of the three small-BZ Fermi-surface pockets shown in Fig.~\ref{fig: smallFS}  to the large BZ. This figure supplements the discussion in Section~\ref{ss:largeBZ}.


\begin{references}

\bibitem{tranquada1995evidence}
J. M. Tranquada et al., Nature {\bf 375}, 561-563 (1995).


\bibitem{yamada1998doping}
K. Yamada et al., Phys. Rev. B {\bf 57}, 6165 (1998).

\bibitem{hoffman2002four}
J.E. Hoffman et al., Science {\bf 295}, 466--469 (2002).

\bibitem{mcelroy2003relating}
K. McElroy et al., Nature {\bf 422}, 592--596 (2003).

\bibitem{vershinin2004local}
M. Vershinin et al., Science {\bf 303}, 1995--1998 (2004).

\bibitem{hanaguri2004checkerboard}
T. Hanaguri et al., Nature {\bf 430}, 1001--1005 (2004).

\bibitem{abbamonte2005spatially}
P. Abbamonte et al., Nat. Phys. {\bf 1}, 155--158 (2005).

\bibitem{mcelroy2005coincidence}
K. McElroy et al., Phys. Rev. Lett. {\bf 94}, 197005 (2005).

\bibitem{wise2008charge}
W.D. Wise et al., Nat. Phys. {\bf 4}, 696--699 (2008).

\bibitem{da2014ubiquitous}
E.H. da Silva Neto et al., Science {\bf 343}, 393--396 (2014).

\bibitem{comin2015symmetry}
R. Comin et al., Nat. Mat. {\bf 14}, 796--800 (2015).



\bibitem{valla2006ground}
T. Valla et al., Science {\bf 314}, 1914-1916 (2006).

\bibitem{he2009energy}
R.-H. He et al., Nat. Phys. {\bf 5},  119-123 (2009).

\bibitem{matt2015electron}
C.E. Matt et al., Phys. Rev. B {\bf 92}, 134524 (2015).

\bibitem{chang2008electronic}
J. Chang et al., New J. Phys. {\bf 10}, 103016 (2008).




\bibitem{barivsic2013universal}
N. Bari{\v{s}}i{\'c} et al., Nat. Phys. {\bf 9}, 761–764 (2013).

\bibitem{doiron2015evidence}
N. Doiron-Leyraud et al., Nat. Commun. {\bf 6}, 6034 (2015).

\bibitem{doiron2007quantum}
N. Doiron-Leyraud et al., Nature {\bf 447}, 565-568 (2007).

\bibitem{sebastian2012quantum}
S.E. Sebastian et al., Phys. Rev. Lett. {\bf 108}, 196403 (2012).

\bibitem{sebastian2009spin}
S.E. Sebastian et al., Phys. Rev. Lett. {\bf 103}, 256405 (2009).

\bibitem{vignolle2008quantum}
B. Vignolle et al., Nature {\bf 455}, 952 (2008).




\bibitem{millis2007antiphase}
A.J. Millis and M.R. Norman, Phys. Rev. B {\bf 76}, 220503 (2007).

\bibitem{chakravarty2008fermi}
S. Chakravarty and H.-Y. Kee, Proc. Natl Acad. Sci. {\bf 105}, 8835--8839 (2008).


\bibitem{chen2008quantum}
W.-Q. Chen, K.-Y. Yang, T.M. Rice and F.C. Zhang, Europhys. Lett. {\bf 82}, 17004 (2008).

\bibitem{zabolotnyy2009evidence}
V. B. Zabolotnyy et al., EPL {\bf 86}, 47005 (2009).


\bibitem{yao2011fermi}
H. Yao, D.-H. Lee and S. Kivelson, Phys. Rev. B {\bf 84}, 012507 (2011).

\bibitem{harrison2011protected}
N. Harrison and S.E. Sebastian, Phys. Rev. Lett. {\bf 106}, 226402 (2011).


\bibitem{allais2014connecting}
A. Allais,	D. Chowdhury, and S. Sachdev, Nat. Commun. {\bf 5}, 5771 (2014).







\bibitem{laliberte2011fermi}
F. Laliberte et al., Nat. Commun. {\bf 2}, 432 (2011).


\bibitem{christensen2007nature}
N.B. Christensen et al., Phys. Rev. Lett. {\bf 98}, 197003 (2007).



\bibitem{fine2004hypothesis}
B.V. Fine, Phys. Rev. B {\bf 70}, 224508 (2004).

\bibitem{fine2007interpretation}
B.V. Fine, Phys. Rev. B {\bf 75}, 014205 (2007).

\bibitem{fine2007magnetic}
B. V. Fine, Phys. Rev. B {\bf 75}, 060504 (2007).

\bibitem{fine2011implications}
B.V. Fine, J. Supercond. Nov. Magn. {\bf 24}, 1207-1211 (2011).



\bibitem{brandenburg2013dimensionality}
J. G. Brandenburg and B.V. Fine,  J. Supercond. Nov. Magn. {\bf 26}, 2621-2626 (2013).



\bibitem{kivelson2003detect}
S.A. Kivelson et al., Rev. Mod. Phys. {\bf 75}, 1201 (2003). 

\bibitem{Robertson2006} J. A. Robertson et al., Phys. Rev. B {\bf 74}, 134507 (2006).



\bibitem{Zaanen1989} J. Zaanen and O. Gunnarsson, Phys. Rev. B {\bf 40}, 7391 (1989).
\bibitem{Seibold2011} G. Seibold, R. S. Markiewicz, and J. Lorenzana, Phys. Rev. B {\bf 83}, 205108 (2011).




\bibitem{SchriefferBook2007}
J. M. Tranquada, Neutron Scattering Studies of Antiferromagnetic correlations in Cuprates, in {\it Handbook of High-Temperature Superconductivity}, Editors: J. R. Schrieffer and  J. S. Brooks, Springer, (2006).
\bibitem{comin2015broken}
R. Comin et al., Science {\bf 347}, 1335--1339 (2015).

\bibitem{fine2016}
B.V. Fine, Science {\bf 351}, 235-a (2016). 

\bibitem{Comin2016}
R. Comin et al.,  Science {\bf 351}, 235-b (2016).

\bibitem{Wang2015} Y. Wang and A. V. Chubukov, Phys. Rev. B {\bf 92}, 245140 (2015).

\bibitem{Jang2016} H. Jang et al., Proc. Natl. Acad. Sci. USA {\bf 113}, 14645 (2016).


\bibitem{Avci2014}  S. Avci et al., Nat. Commun. {\bf 5}, 3845 (2014).
\bibitem{Bohmer2015} A.E. B\"ohmer et al., Nat. Commun. {\bf 6}, 7911 (2015).
\bibitem{Ohalloran2017} J. O'Halloran et al., Phys. Rev. B {\bf 95}, 075104 (2017).
\bibitem{Meier2017} W. R. Meier et al., eprint arXiv:1706.01067 (2017).


\bibitem{geim2007rise}
A.K. Geim and K.S. Novoselov, Nat. Mat. {\bf 6}, 183-191 (2007).



\bibitem{pavarini2001band}
E. Pavarini et al., Phys. Rev. Lett. {\bf 87}, 047003 (2001).


\bibitem{kojima2000magnetism}
K.M. Kojima et al., Physica B {\bf 289},  343-346 (2000).



\bibitem{egami2010spin}
T. Egami et al., Adv. Cond. Matter. Phys. {\bf 2010} (2010).

\bibitem{ashcroft1976solid}
N.W. Ashcroft and N.D. Mermin, Solid state physics, Saunders College, Philadelphia (1976). 



\bibitem{li2007two}
Q. Li et al., Phys. Rev. Lett. {\bf 99}, 067001 (2007).

\bibitem{tranquada2008evidence}
J.M. Tranquada et al., Phys. Rev. B {\bf 78}, 174529 (2008).

\bibitem{Berg2007} E. Berg et al., Phys. Rev. Lett. {\bf 99}, 127003 (2007).

\bibitem{Bhartiya2017} V. Bhartiya and B. V. Fine, eprint arXiv:1703.09979 (2017).






\end{references}
\end{document}